# Bayesian inference over model-spaces increases the accuracy of model comparison and allows formal testing of hypotheses about model distributions in experimental populations.

Thomas HB FitzGerald[1,2,3], Dorothea Hämmerer[2,4,5,6], Thomas D. Sambrook[1], Will D. Penny[1,2]

1. School of Psychology, University of East Anglia, Norwich Research Park, Norwich, Norfolk, NR4 7TJ, UK
2. The Wellcome Trust Centre for Neuroimaging, University College London, 12 Queen Square, London, WC1N 3BG, UK
3. Max Planck-UCL Centre for Computational Psychiatry and Ageing Research, Russell Square House, London, WC1B 5EH, UK
4. Chair of Lifespan Developmental Neuroscience, Department of Psychology, TU 16 Dresden, Zellescher Weg 17, 01069 Dresden, Germany
5. Institute of Cognitive Neuroscience, University College London, 17 Queen Square, 18 London, WC1N3AZ, UK
6. Institute of Cognitive Neurology and Dementia Research, Otto-von-Guericke 20 University Magdeburg, Leipziger Str. 44, 39120 Magdeburg, Germany

Correspondence: Thomas FitzGerald

School of Psychology, University of East Anglia, Norwich Research Park, Norwich, Norfolk, NR4 7TJ, UK

E-mail: t.fitzgerald@uea.ac.uk.

Telephone: +44 1603 59 2270

**Abstract**

Determining the best model or models for a particular data set, a process known as Bayesian model comparison, is a critical part of probabilistic inference. Typically, this process assumes a fixed model-space (that is, a fixed set of candidate models). However, it is also possible to perform Bayesian inference over model-spaces themselves, thus determining which spaces provide the best explanation for observed data. Model-space inference (MSI) allows the effective exclusion of poorly performing models (a process analogous to Automatic Relevance Detection), and thus mitigates against the well-known phenomenon of model dilution, resulting in posterior probability estimates that are, on average, more accurate than those produced when using a fixed model-space. We focus on model comparison in the context of multiple independent data sets (as produced, for example, by multi-subject behavioural or neuroimaging studies), and cast our proposal as a development of random-effects Bayesian Model Selection, the current state-of-the-art in the field. We demonstrate the increased accuracy of MSI using simulated behavioural and neuroimaging data, as well as by assessing predictive performance in previously-acquired empirical data. Additionally, we explore other applications of MSI, including formal testing for a diversity of models within a population, and comparison of model-spaces between populations. Our approach thus provides an important new tool for model comparison.

**Introduction**

A key step in any experimental study is that of determining which hypotheses to test using the data that have been acquired. From a Bayesian perspective, this corresponds to choosing which models to fit to the data, with each model constituting a distinct hypothesis about how the data were generated. An experimenter is thus faced with the problem of choosing a model-space (or set of models) that is appropriate to address their questions. In large part, this depends upon scientific creativity, in concert with knowledge of preceding literature. However, where a number of models are of interest, the phenomenon of model dilution is encountered [1,2]. Here, the posterior probability assigned to each model, including the 'true' or 'best' model, progressively decreases with model-space size. This occurs because each model is assigned non-zero prior probability (typically the prior over models will be uniform), and a non-zero likelihood for each data set being considered. Since probabilities must sum to one, this results in overly conservative posteriors. (See below for a fuller discussion of this in the context of multi-subject studies, where this problem is particularly significant)

Model dilution presents experimenters with a dilemma, whether to arbitrarily exclude models of potential importance from consideration, or to accept conservatively-biased results. To mitigate against this, we propose the additional step of inferring over model-spaces (which we will term model-space inference: MSI). This allows consideration of situations where a particular model is excluded from the model-space (and thus assigned zero probability), resulting in more accurate posterior probability estimates. (This is demonstrated below). The capacity of MSI to 'prune' models is analogous to Automatic Relevance Determination (ARD) [3,4]. In ARD one effectively discards parameters that are not needed to explain the data. MSI allows one to discard models in a similar way, thus allows consideration of a broader hypothesis space whilst maintaining sensitivity.

Model-space inference takes on an additional dimension in the context of multiple independent data sets, each of which may have been generated by a different process. (This is generally the case, for example, where one has data from multiple experimental subjects) Here, inference over model-spaces corresponds to inferring on the distribution of models within an experimental population. For example, one might have a situation in which different subjects solved a particular cognitive task using two or more different strategies [5,6]. In addition to providing a more accurate characterisation of this distribution, MSI provides a formal framework for addressing questions like whether there is a genuine diversity of models within a population, or whether model-distributions are similar in different populations [7]. (For example, whether there is a similar distribution of learning strategies in patients and in healthy controls)

In this paper, we focus on implementing model-space inference in the context of data acquired from multiple subjects. Here, MSI constitutes an extension to, and improvement of, random effects Bayesian model selection (rBMS) [7,8], an approach to model comparison that is widely used in neuroscientific and behavioural data analysis, as well as other areas. The methods presented here thus have the potential to increase the accuracy of data analysis across a range of studies and methodologies.

**Methods & Materials**

**Bayesian model comparison**

In model comparison problems, one wishes to assign posterior probabilities to a set of candidate models $m_1$ to $m_I$, each of which constitutes a distinct hypothesis, based on one or more data sets $y_1$ to $y_N$. In the simplest case, where there is only a single data set $y$, it is straightforward to calculate the Bayes Factor for each model based on the log model evidences $\ln p(y|m_i)$ (or some approximation to them) and the prior probability of each model obtaining $p(m_i)$:

$$\ln p(m_i | y) \propto \ln p(y_n | m_i) + \ln p(m_i). \tag{1}$$

Where multiple independent data sets are available (for example, where data have been collected from different subjects), this approach can be extended to give the Group Bayes Factor [8,9]:

$$\ln p(m_i | y_1,\ldots,y_N) \propto \sum_{n=1}^{N} \ln p(y_n | m_i) + \ln p(m_i). \tag{2}$$

However, this approach makes the assumption that a single model provides the best explanation for every data set, an assumption that is likely to be unwarranted in many cases [7,8].

An alternative approach, which has become widely used, embeds the model-space of interest within a hierarchical group-level model. This approach, known as random effects Bayesian model selection (rBMS) [7,8], allows for a diversity of models within the sample. Here, both the group-level distribution over models, and the distributions for each particular subject, are estimated simultaneously. (For a formal treatment of this, see [8] and the 'Bayesian Inference over Model-Spaces' section of this paper. See also [10] for a recent development of rBMS to include hierarchical parameter estimation)

Critically, however, all three of these approaches make use of a single model-space, and thus are prone to being 'conservative' (that is, to underestimate the probability of likely models, and overestimate the probability of unlikely ones). In what follows, we focus on model-space inference as an extension of rBMS, both for simplicity, and because it is the application that we envisage having the greatest practical benefit to researchers. However, most of the key points apply equally, whichever method is employed.

**Bayesian inference over model-spaces**

We consider a situation in which we have a model-space comprising of a totally ordered set of $I$ candidate models $\lambda = \{m_1, \cdots, m_I\}$, and $N$ independent datasets $\mathbf{Y} = \{\mathbf{y}_1,\ldots,\mathbf{y}_N\}$ to which they will be applied (Figure 1). Building on previous work [8], we define a set of $J$ hierarchical models, or *metamodels*, each of which corresponds to a different model-space. Each metamodel $S_j$ is a 2-tuple $(\kappa_j, \boldsymbol{\alpha}_{\bullet j}^{prior})$, where $\kappa_j \subseteq \lambda$ is a subset of the model space of size $K_j = |\kappa_j|$, and $\boldsymbol{\alpha}_{\bullet j}^{prior} = \left[\alpha_{1j}^{prior},\ldots,\alpha_{K_j j}^{prior}\right]$ is a set of concentration parameters governing the Dirichlet prior distribution over subject-level models. Thus metamodel $j$ considers a model-space containing $K_j$ models. If a model is excluded from the model-space, then it is assigned zero prior probability. (In previous work, prior concentration parameters are typically all set to one [7,8,10], but this need not be the case [2]). Each metamodel can now be specified fully as described in [2,8] Thus, the variable $\mathbf{r} = \left[r_1,\ldots,r_{K_j}\right]$ assigned to each subject-level model indicates the frequency with which that model is used in the subject population. This is specified as follows using Dirichlet distributions with concentration parameters $\boldsymbol{\alpha}_{\bullet j}$.

$$p(r \mid S_j) = Dir(r_k, \boldsymbol{\alpha}_{\bullet j}) = \frac{1}{Z(\boldsymbol{\alpha}_{\bullet j})} r_k^{\alpha_{kj}-1},$$

$$Z(\boldsymbol{\alpha}_{\bullet j}) = \frac{\sum_{k=1}^{K_j} \Gamma(\alpha_{kj})}{\Gamma(\sum \boldsymbol{\alpha}_{\bullet j})}$$

(3)

where $\Gamma$ indicates the gamma function. These probabilities parameterise separate multinomial distributions for each subject, such that:

$$p(\mathbf{a}_{\bullet n} \mid \mathbf{r}) = \prod_{k=1}^{K_j} r_k^{a_{nk}},$$

(4)

where $\mathbf{A}$ is an $N \times K_j$ matrix of indicator variables for all subjects and models included within the model space, such that $a_{nk} \in \{0,1\}$ and $\forall_n : \sum \mathbf{a}_{n\bullet} = 1$.

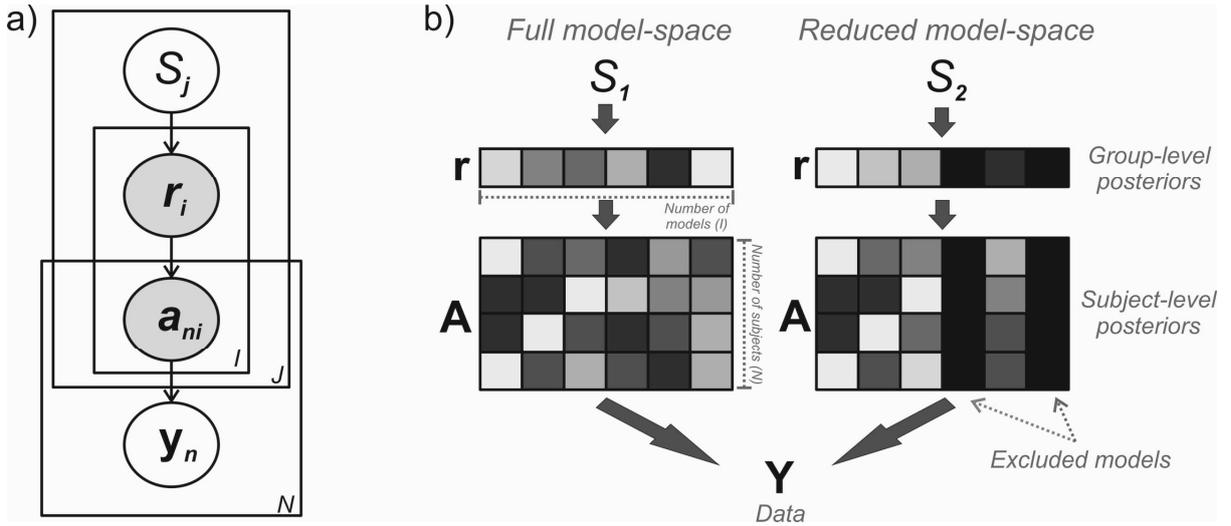

Figure 1: Illustration of metamodel structure. Figure 1a shows the metamodel structure formally, as a directed acyclic graph. (Filled circles indicate unknown variables). Figure 1b provides a cartoon illustration of the difference between metamodels containing the full model-space ($S_1$), in which both group-level and subject-level posteriors are estimated for all subjects and models, and a reduced model-space ($S_2$), in which models 3 and 5 have been excluded from the model space (assigned zero probability). (The colour of each cell corresponds to the (cartoon) estimated posterior probabilities, with white indicating a probability of one, and black indicating a probability of zero)

Thus, after marginalising out model parameters $\vartheta$ the likelihood is given by :

$$p(\mathbf{Y} \mid \mathbf{A}) = \exp\left(\sum_{n=1}^{N} \sum_{k=1}^{K_j} a_{nk} l_{nk}\right),$$

(5)

where $l_{nk} \approx \ln p(\mathbf{y}_n | a_{nk} = 1)$ is the approximate (or in some cases true) log model evidence for model $k$ in subject $n$ [7]. (Typically, estimating this is performed separately from model selection. However, in principle these steps can be integrated, which has considerable attraction in the context of hierarchical parameter estimation [10])

The full joint probability distribution now becomes

$$p(\mathbf{Y}, \mathbf{A}, \mathbf{r} | S_j) = p(\mathbf{y} | \mathbf{A}) p(\mathbf{A} | \mathbf{r}) p(\mathbf{r} | S_j),$$
$$= \exp\left(\sum_{n=1}^{N} \sum_{k=1}^{K_j} a_{nk} l_{nk}\right) \prod_{k=1}^{K_j} r_k^{a_{nk}} \frac{1}{Z(\boldsymbol{\alpha}_{\bullet j})} \prod_{k=1}^{K_j} r_k^{\alpha_{kj}-1}. \tag{6}$$

(This structure is depicted in Figure 1)

*Calculating the metamodel evidence using variational Bayes*

The posterior probability for each metamodel is given by

$$p(S | \mathbf{Y}) \propto p(\mathbf{Y} | S) p(S), \tag{7}$$

Thus, under the assumption that the prior distribution over metamodels $p(S)$ is known (this is assumed to be uniform, unless otherwise stated), all that is needed need to compute this quantity is the metamodel evidence $p(\mathbf{Y} | S)$. Given the hierarchical model structure described above and in [8], this becomes:

$$p(\mathbf{Y} | S) = \sum_{j=1}^{J} \sum_{n=1}^{N} \sum_{k=1}^{K_j} \int p(\mathbf{Y} | a_{nk} = 1) p(a_{nk} = 1 | \mathbf{r}) p(\mathbf{r} | S_j) dr. \tag{8}$$

Exact evaluation of the metamodel evidence is impossible, but under the variational scheme proposed in [8], a variational lower bound is easy to calculate. First, inference is performed using variational Bayes [11], which can be done simply by iterating the following equations until convergence, making use of the fact that $l_{nk} = \ln p(\mathbf{y}_n | a_{nk} = 1)$ has been precomputed:

$$u_{nk} = \exp\left(l_{nk} + \Psi(\alpha_{kj}) - \Psi\left(\sum \boldsymbol{\alpha}_{\bullet j}\right)\right),$$
$$g_{nk} = p(a_{nk} = 1 | \mathbf{Y}) = \frac{u_{nk}}{\sum \mathbf{u}_{n\bullet}}, \tag{9}$$
$$\alpha_{kj} = \alpha_{kj}^{prior} + \sum_{n=1}^{N} g_{nk},$$

where $\Psi$ indicates the digamma function. This yields a variational approximation to the posterior $q(\mathbf{A}, \mathbf{r}) = q(\mathbf{A}) q(\mathbf{r})$ over the metamodel parameters, and the variational lower bound (also known as the negative variational free energy) $L_j$ for the $j$-th metamodel, is given by:

$$L_j = E_{m,r}\left[\ln p(\mathbf{Y} | \mathbf{A}) + \ln p(\mathbf{A} | \mathbf{r}) + \ln p(\mathbf{r} | S_j) - \ln q(\mathbf{A}) - \ln q(\mathbf{r})\right], \tag{10}$$

This is identical to the evidence equation in Rigoux et al. [7] for the 'complete' metamodel (in other words, the one that contains the whole model-space). Here we generalise these equations to accommodate reduced model spaces (incomplete metamodels). Thus, using *(3)*, *(4)* and *(5)*:

$$E_{m,r}\left[\ln p(\mathbf{Y}|\mathbf{A})\right] = \sum_{n=1}^{N}\sum_{k=1}^{K_j} g_{nk} l_{nk},$$

$$E_{m,r}\left[\ln p(\mathbf{A}|\mathbf{r})\right] = \sum_{n=1}^{N}\sum_{k=1}^{K_j} g_{ni}\left(\Psi(\alpha_{kj}) - \Psi(\sum \boldsymbol{\alpha}_{\bullet j})\right),$$

$$E_{m,r}\left[\ln p(\mathbf{r}|S_j)\right] = \sum_{k=1}^{K_j}(\alpha_{0ij}-1)\left(\Psi(\alpha_{kj}) - \Psi(\sum \boldsymbol{\alpha}_{\bullet j})\right) - \ln Z(\boldsymbol{\alpha}_{\bullet j}^{prior}), \quad (11)$$

$$E_{m,r}\left[\ln q(\mathbf{A})\right] = \sum_{n=1}^{N}\sum_{k=1}^{K_j} g_{nk} \ln g_{nk},$$

$$E_{m,r}\left[\ln q(\mathbf{r})\right] = \sum_{k=1}^{K_j}\ln \Gamma(\alpha_{kj}) - \ln \Gamma(\sum \boldsymbol{\alpha}_{\bullet j}) - \sum_{k=1}^{K_j}(\alpha_{kj}-1)\left(\Psi(\alpha_{ki}) - \Psi(\sum \boldsymbol{\alpha}_{\bullet j})\right).$$

$L_j$ can now be used for comparison between different model-spaces. (Note that, although we focus on this variational scheme here, the principles of model-space comparison we describe are equally applicable to alternative inference schemes, so long as they provide an approximation to the metamodel evidence)

*Model-space comparison*

Having calculated an approximate model evidence for each models-space, we can now compare different spaces using the standard principles of Bayesian model comparison [1,12,13] Thus:

$$\hat{R}_j = p(S_j|\mathbf{Y}) = \frac{e^{L_j} U_j}{\sum_{j=1}^{J} e^{L_j} U_j}, \quad (12)$$

where $U_j = p(S_j = 1)$ is the prior probability distribution over models-spaces, which we will assume to be uniform unless otherwise stated. Having calculated the posterior probability over model-spaces, one can now either select the best one (model-space selection: MSS), or average over model-spaces (model-space averaging: MSA) [1,14]. We focus on this latter approach, which is strictly optimal, although MSS can be thought of as a limiting case of MSA [15].

The set of model-spaces to be considered can be defined in a number of ways. The simplest approach is to consider all possible (full and reduced) models-spaces (an exhaustive search), or alternatively some subset specified *a priori*. A complementary approach, of potential utility where the number of models is very large, or model-space calculations slow [10], is to perform a greedy search over model-spaces. Here we first eliminate the model with the lowest posterior probability in the full model-space, then recalculate the metamodel evidence. If this evidence has increased, we then remove the model with the lowest posterior probability from the new metamodel and recalculate. If not, we try removing the model with the second lowest posterior probability from the complete metamodel, and so on. This process is iterated until all models within the current 'best' model-space have been tried, without further improving the metamodel evidence.

*Calculating posterior and exceedance probabilities using MSA*

Posterior probabilities over subject-level models conditioned on model-space $j$, $\hat{\mathbf{r}}_{\bullet j} = \left[ \hat{r}_{1j}, \ldots, \hat{r}_{Ij} \right]$, can be calculated using the expected values of the multinomial parameters as follows [8]:

$$\hat{r}_{kj} = \begin{cases} \alpha_{ij} / \sum \boldsymbol{\alpha}_{\bullet j} & \text{if } m_k \in \kappa_j \\ 0 & \text{if } m_k \notin \kappa_j \end{cases}, \quad (13)$$

where $i : \kappa_{ij} = m_k$ is used to index those models present in the *j*-th model-space. It is now straightforward to calculate the posterior probabilities averaged across model-spaces, using the principles of Bayesian model averaging [1,13]:

$$r'_k = \sum_{j=1}^{J} \hat{r}_{kj} R_j. \quad (14)$$

Alternatively, under MSS we simply use the posteriors from the best model-space:

$$\hat{r}_k^{MSS} = \hat{r}_{kj} \quad : j = \arg\max_j \left( p(S \mid \mathbf{Y}) \right). \quad (15)$$

Calculating exceedance probabilities (the belief that one model is more likely than any other model: $p(r_k \geq r_{k \neq k} \mid \mathbf{Y})$) during model-space averaging, can be performed in a similar fashion using the exceedance probabilities $\boldsymbol{\varphi}_{\bullet j} = \left[ \varphi_{1j}, \ldots, \varphi_{Ij} \right]$ from individual metamodels (calculated as described in [8]:

$$\varphi'_k = \sum_{j=1}^{J} \varphi_{kj} P_j,$$

$$\varphi_k^{MSS} = \varphi_{kj} \quad : j = \arg\max_j \left( p(S \mid \mathbf{Y}) \right). \quad (16)$$

However, exceedance probabilities are known to be overconfident, making it preferable to use protected exceedance probabilities, as described in [7]. To calculate these, we first find the approximate model evidence $F_0$ for a 'null' metamodel $S_0$, as described by [7]. (Here 'null' means a metamodel in which all models are fixed as having the same probability) This is then used to calculate the 'protected posterior probability' $\tilde{R}_j$ of each metamodel within the expanded metamodel space, using a prior distribution over metamodels such that the null metamodel has a prior probability of 0.5, whilst those of all other metamodels are equal to one another. Thus:

$$\tilde{R}_j = \frac{e^{L_j + \ln(U_j)}}{\sum_{j=0}^{J} e^{L_j + \ln(U_j)}}$$

$$U_j = \begin{cases} 0.5 J^{-1} & \text{if } j \in \{1, \ldots, J\} \\ 0.5 & \text{if } j = 0 \end{cases}, \quad (17)$$

$$\sum_{j=0}^{J} U_j = 1.$$

The protected exceedance probabilities can now be calculated as:

$$\tilde{\varphi}'_k = \sum_{j=0}^{J} \varphi_{kj} \tilde{R}_j. \quad (18)$$

**Practical applications of model-space comparison**

*Application 1: Model-space averaging reduces the conservative bias inherent in rBMS*

Random effects BMS makes the implicit assumption that all models within the model-space have a non-zero probability of generating the data, as reflected in the prior parameters of the Dirichlet distribution. This means that, assuming that the distribution is symmetrical with $\alpha_{kj}^{prior} = c$ for all values of $k$, the minimum posterior probability that can be assigned to any model is $\tau := c/(n+kc)$, and, rather than lying in the range $1 \geq \hat{r}_{kj} \geq 0$, the posterior probabilities are constrained to lie within the range $1-(n-1)\tau > \hat{r}_{kj} > \tau$. Thus only where no models have a true probability outside this range can rBMS accurately estimate the posterior probabilities, even in principle. More generally, the fact that rBMS cannot reduce the posterior probability assigned to any subject-level model below $\tau$, however poorly that model explains the data, means that it tends, in general, to overestimate the posterior probabilities of unlikely models, and underestimate the posterior probabilities of likely ones. It can thus be thought of as being conservative, and the effects of this can be seen in Figures 1 and 2.

MSA, by contrast, permits the (effective) elimination of models. This occurs when all model-spaces in which they are included are assigned a posterior probability of close to zero, and can be thought of by analogy with Automatic Relevance Detection [3,4], which effectively eliminates redundant parameters by assigning them a value very close to zero. This capacity to eliminate redundant models largely removes the conservative bias inherent in rBMS, and allows posterior probabilities to take the range $0 < r'_k < 1$.

Conversely, where the full model-space applies and rBMS is an optimal strategy, MSA will tend to closely approximate rBMS, and any differences between the results of these processes are likely to be small. This occurs because, in these situations, the posterior probability assigned to the complete metamodel will tend towards one, and illustrates the fact that rBMS can be thought of as a limiting case of MSA (where $U_1 = 1$ and $U_{j>1} = 0$). Thus overall, MSA has the potential to significantly increase the accuracy of model comparison when compared with rBMS, particularly in situations where the data are best explained by a small subset of the full model-space, or $n$ is small.

*Application 2: Testing for population diversity*

In many situations model comparison, however performed, leads to results where significant posterior probabilities are assigned to more than one model (Table 1). In such situations, it is important to know whether these results reflect true population diversity, or simply inconclusive data. For example, one might wish to know whether a dataset supported the employment of two different cognitive strategies by different subjects, or whether, by contrast, it simply is not possible based on that data set to infer which strategy is being used. This can be formally addressed using model-space comparison, by comparing the evidence for model-spaces containing only one of the models, with the evidence for the model-space containing both. (For simplicity, we focus on the simplest case, in which only two models are assigned a significant probability, but the same principles generalise to more complex cases) Model-space comparison thus provides a potentially important tool for quantifying the evidence for diversity within a population.

| | Dataset 1: Two-step task [19] | | | | |
|---|---|---|---|---|---|
| | Hybrid learner | Pure Model-free learner | Pure Model-based learner | | |
| rBMS ($\hat{r}$) | 0.51 | 0.02 | 0.47 | | |
| MSA ($r'$) | 0.52 | 0.01 | 0.48 | | |
| | Dataset 2: Probabilistic inference task [5] | | | | |
| | Sequence length = 1 | Sequence length = 2 | Sequence length = 3 | Sequence length = 4 | Sequence length = 5 |
| rBMS ($\hat{r}$) | 0.05 | 0.62 | 0.25 | 0.04 | 0.04 |
| MSA ($r'$) | 0.01 | 0.66 | 0.31 | 0.01 | 0.01 |

**Table 1: Estimated posterior probabilities for two previously acquired datasets using rBMS and MSA. In keeping with both our theoretical arguments and simulations, the results of rBMS show evidence of being conservative, when compared with those of MSA. (The probabilities assigned to likely models are lower, and those assigned to unlikely models are higher) These differences are fairly small, reflecting the relatively low number of models and high number of subjects in both studies, and can be expected to be greater in many other applications.**

*Application 3: Comparing populations using model-space comparison*

Model-space comparison can also be used to test whether the distribution of models in different experimental populations is the same or not. (This could be used, for example, to test for differences between age groups or nationalities) Thus, for $T$ populations, each of which contains subjects $w_t \subset N$, we can compare the approximate model evidences under the hypotheses of a single ($L_0^w$) or multiple identically-specified ($L_1^w$) metamodels

$$\begin{aligned}
L_0^w &\approx \ln p(\mathbf{Y} | S_1), \\
L_1^w &\approx \sum_{j=1:T} \ln p(\mathbf{Y}_{w_t} | S_j), \\
&\{\kappa_j : \forall \kappa_j, \kappa_j = \lambda\}, \\
&\{\boldsymbol{\alpha}_{:j}^{prior} : \forall_j \forall_k, \alpha_{kj}^{prior} = c\},
\end{aligned} \quad (19)$$

where $c > 0$ is a constant. This tests whether the evidence is greater when all subjects are hypothesised to be from the same population, compared with when they are hypothesised to come from different populations. (It can thus be thought of by analogy with an ANOVA over populations in classical statistics, where one tests for the main effect of population.) This form of analysis has previously been proposed by Rigoux et al. [7]. Our key contribution is to situate it within the broader framework of model-space inference, of which it forms a special case.

## Simulations

*Simulating behaviour on a learning task*

We simulated behaviour using simple reinforcement learning agents on a two-arm bandit task with drifting reward probabilities. These probabilities were generated using a zero-mean Gaussian random walk in logit space, with a variance of 0.5, starting values of logit(0.3) and logit(0.7) for the two bandits, and reflecting boundaries and logit(0.1) and logit (0.9). A single 120 trial instantiation of the task contingencies was used for all simulations reported here (Figure 2a), though the precise feedback given to each simulated agent differed depending on their behaviour.

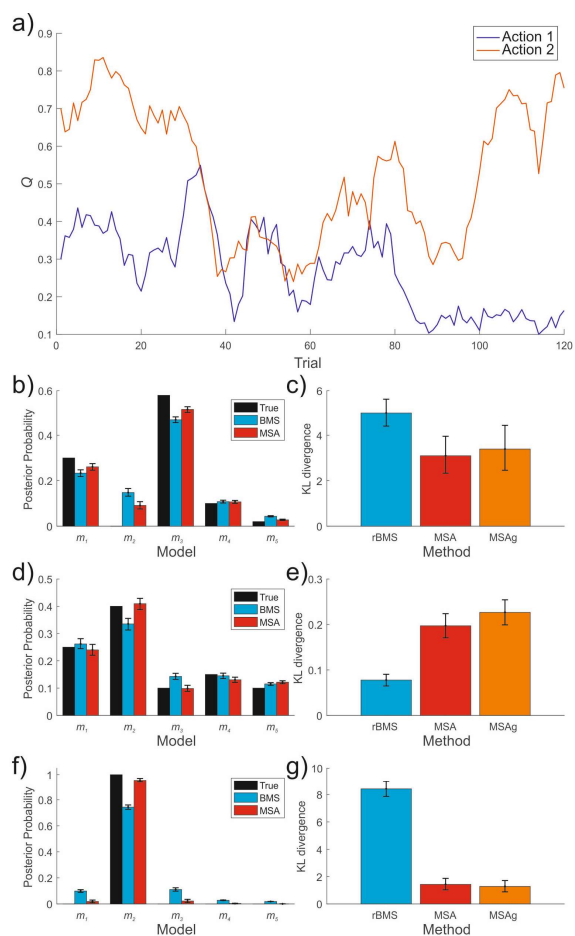

**Figure 2: Comparison of the accuracy of random-effects Bayesian model selection and model-space averaging using simulated behavioural data on a reward learning task. 2a: Data were simulated using reward probabilities ('*Q*') as depicted here, when the agent chose action one (blue) and two (orange). Three different (true) model spaces were simulated (see Methods for more detail). In the first (Top row), we simulated a diverse population, in which one model was absent, and another present only in a single subject. In the second, we simulated a diverse population, in which each model occurred with a probability well above zero In the third, the data were generated using only a single model ($m_2$). The left column shows true probabilities (black) and posterior probabilities generated by rBMS (blue) and MSA using an exhaustive search (red). The right column shows average Kullback-Leibler divergence values for rBMS (blue), MSA using an exhaustive search (labelled 'MSA', red), and MSA using a greedy search (labelled 'MSAg', orange). Where one or models had a zero occurrence probability (2b,c and 2f,g) MSA produced**

**considerably more accurate results, particularly in the case where the data were generated by a single model (bottom row). In context two (2d,e) rBMS was slightly more accurate, reflecting the close match between the structure of the full metamodel it uses and the true distribution. However, the magnitude of these differences was small (note the different scales in 3c,e,g), reflecting the capacity of MSA to approximate rBMS where appropriate. Encouragingly, use of a greedy search during MSA resulted in similar levels of accuracy to those found when using an exhaustive search, supporting the use of a greedy search when motivated by computational demands. (Error bars indicate bootstrapped 95% confidence intervals)**

The simulated agents employed simple Q-learning [16,17], meaning that on each trial beliefs about action values were updated according to the following equation:

$$Q(z_t) = Q(z_t) + \eta(y_t - Q(z_t)) \tag{20}$$

Where $z_t$ is the bandit chosen at trial $t$, $y_t$ is the binary outcome (reward or no reward), and $\eta$ is the learning rate.

Action probabilities were generated using a softmax decision rule:

$$p(z_t = i) = \frac{\exp(\beta Q(z_i))}{\sum_{i=1}^{2} \exp(\beta Q(z_i))} \tag{21}$$

The $\beta$ (inverse temperature) parameter, which indexes behavioural stochasticity was varied between simulations, as described in the text. In all cases, action values were initialised at 0.5.

*Comparing the performance of rBMS and MSA using simulated behavioural data*

We next considered the more general problem of assigning posterior probabilities to subject-level models. To do so, we simulated behaviour on the reward learning task described above using five agent models. Models one to four were simple *Q*-learning agents, differentiated by their learning rates ($m_1$: $\eta = 0.2$. $m_2$: $\eta = 0.5$. $m_3$: $\eta = 0.8$. $m_4$: $\eta = 0$) (note that a learning rate of zero corresponds to action selection at chance). All four agents used a softmax decision rule with an identical inverse temperature parameter ($\beta = 1$). Agent five ($m_5$) was identical to agent 1, except that action probabilities were inverted (in other words, the agent tended to choose what it believed to be the lower value option). 50 subjects were simulated, and the resulting data used to carry out random effects BMS [8], and MSA using both exhaustive and greedy searches. We will refer to exhaustive search as MSA and greedy search as MSAg for clarity. (In practice, these data would more conventionally be analysed by fitting the learning rate and inverse temperature as free parameters, but we have selected these simulations as a minimally complex way to illustrate our key points). Three contexts were simulated, corresponding to different underlying distributions of subject-level models. The distribution for context 1 was: [$m_1$: 0.3, $m_2$: 0, $m_3$: 0.58, $m_4$: 0.1, $m_5$: 0.02], for context 2: [$m_1$: 0.25, $m_2$: 0.4, $m_3$: 0.1, $m_4$: 0.15, $m_5$: 0.1], ad for context 3: [$m_1$: 0, $m_2$: 1, $m_3$: 0, $m_4$: 0, $m_5$: 0]. These were designed to reflect a range of possible situations, the first a situation in which several of the candidate models truly generated the data, but at least one had an occurrence probability of zero. The second, a situation in which there was a broad spread of probability across models, and all those considered contributed significantly to the data. The third, a situation in which only one of the models generated the data. The simulations are not meant to be exhaustive, but to give a broad idea of how these different analysis strategies may be expected to perform.

*Dissociating population diversity from inconclusive data*

To illustrate the utility of model-space comparison for quantifying the evidence for population diversity, we simulated behaviour on a two-arm bandit task in two contexts. In context one (the 'diverse population' context), half (40) of the experimental data sets were generated using an agent with a learning rate of $\eta = 0.2$, and half with a learning rate of $\eta = 0.8$. For both agents, action selection was performed using a softmax decision-rule with an inverse temperature of $\beta = 8$ (in other words, subjects chose the option that they believed to be best close to deterministically). In context two, the 'inconclusive data' context, all (80) of the data sets were generated using an agent with $\eta = 0.2$, but this time with $\beta = 0.01$, making behaviour extremely stochastic. For each simulated data set, we calculated the log likelihood of the observed data for each of four models (note that since there were no free parameters, this corresponds to the model evidence). ($m_1$: $\eta = 0.2$, $\beta = 8$. $m_2$: $\eta = 0.8$, $\beta = 8$. $m_3$: $\eta = 0.2$, $\beta = 0.01$. $m_4$: $\eta = 0.8$, $\beta = 0.01$.) These simulations were performed 100 times, and the posterior probabilities for the full metamodel ($S_1$) calculated each time. Additionally, we performed model-space comparison between the full metamodel and a set of reduced model-spaces (see Figure 3a).

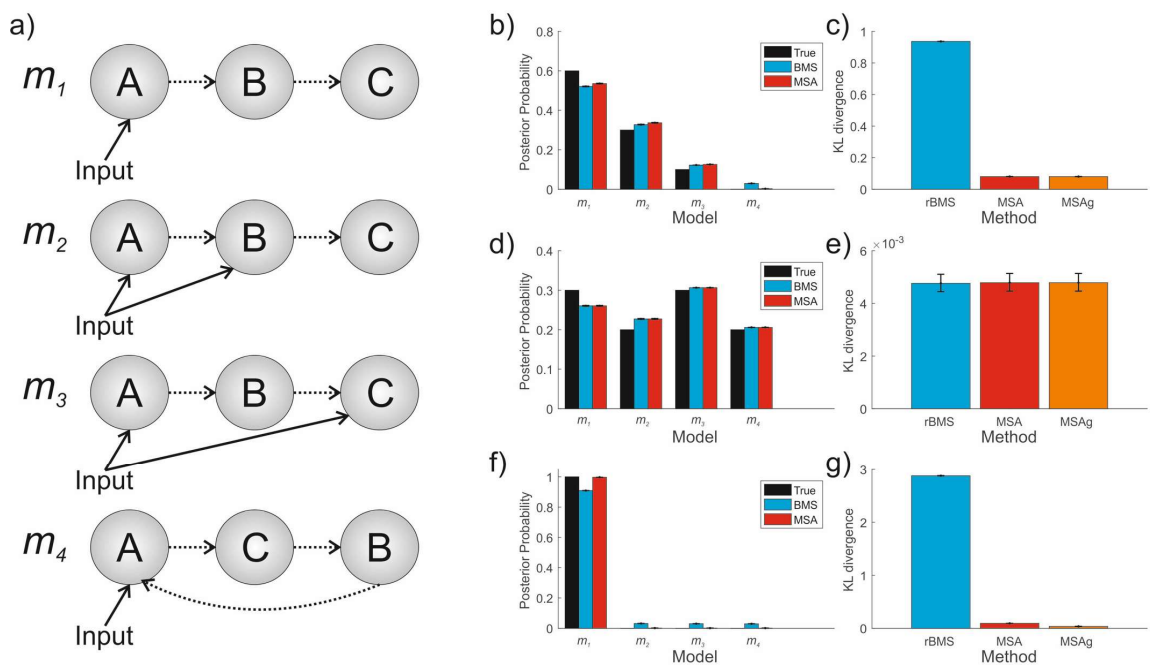

**Figure 3: Comparison of random effects BMS and MSA using simulated neuroimaging data. Data were simulated using four different models (3a), and these models were then fitted to the simulated time series. Model-space averaging produced posterior probability estimates that were significantly more accurate than those produced by rBMS in contexts where one or more models was very unlikely (context one: 3b, 3c and context 3: 3f, 3g), as revealed by both posterior mean posteriors (middle column) and the mean Kullback-Leibler divergence between estimated and true distributions (right column). Where all models were fairly likely (context two: 3d, 3e) the results of both approaches showed comparable accuracy. ('MSA' indicates model-space averaging using an**

exhaustive search. 'MSAg' indicates use of a greedy search. Error bars indicate bootstrapped 95% confidence intervals)

*Comparing population distributions using simulated behavioural data*

To illustrate the use of model-space comparison to test for differences between the distributions of models in different populations, we simulated the reward learning task as described above with two samples of fifty subjects (these might correspond, in a real experiment, to younger and older adults, or patient and control groups). In context one (the 'separate distibutions' context), population one had the distribution [$m_1$: 0.2, $m_2$: 0.6, $m_3$: 0.1, $m_4$: 0, $m_5$: 0.1] and population two had the distribution [$m_1$: 0.6, $m_2$: 0.2, $m_3$: 0, $m_4$: 0.15, $m_5$: 0.05]. In context two (the 'single distribution' context), both populations had the distribution [$m_1$: 0.2, $m_2$: 0.6, $m_3$: 0.1, $m_4$: 0, $m_5$: 0.1]. Each context was simulated 100 times, and the single distribution ($L_0^w$) and separate distribution ($L_1^w$) calculated as described above.

*Comparing the performance of rBMS and MSA using simulated neuroimaging data*

We also compared the performance of rBMS and MSA using dynamic causal modelling (DCM) of simulated fMRI data [18], performed using SPM12 (https://www.fil.ion.ucl.ac.uk/spm/). Data were simulated from three regions, using a single exogenous input to region one, and a model space of possible connectivity patterns as depicted in Figure 3a. Inputs consisted of a series of delta functions spaced 10 seconds apart, and data were simulated from 100 scans, acquired with spacing of 3 seconds. Input and connection strengths were left at their default values, the SPM canonical haemodynamic response function was used, and a signal-to-noise ratio of 1 was employed. (Simulations were performed using the function *spm_dcm_create*). Both analyses and simulations were performed using deterministic, single state models, and the negative variational free energy was used as an approximation to the log model evidence.

**Empirical data**

*Description of datasets*

To explore the performance of model-space comparison using real data, we used the results of two previously published studies. We briefly describe these here, but for fuller details readers are referred to the original papers. In both cases, exhaustive searches over metamodel space were performed.

Dataset 1: Behavioural results, in the form of Bayesian Information Criterion (BIC) scores reported in [19], using the popular two-step task [6]. Briefly, three models were fitted to data from 45 subjects, a pure 'model-based' reinforcement learner, a pure 'model-free' learner, and a hybrid model [6,19]. The study was approved by the ethics committee of the Faculty of Health and Human Sciences at the University of Plymouth.

Dataset 2: Behavioural results from a study looking at inference over sequences of states [5], in the form of BIC scores. Five models were fitted to data collected from 79 subjects, corresponding to inference over sequences of length one to five, respectively. Of these, 43 subjects were younger adults (mean age = 26.4 years, range = 20-33 years) and 36 older adults (mean age = 66,4 years, range = 60-73 years). The study was approved by the local ethics committee of the Charité, University Medicine Berlin.

*Assessing predictive performance using cross-validation*

For real data, the true generative processes are unknown, so to compare the performance of MSA and rBMS in this context, we assessed predictive accuracy using leaveone-out cross validation. To do so, we first separate the data into training data (that from all subjects other than subject $n$ ) and test data (that from subject $n$). (Thus $\mathbf{Y} = \{\mathbf{y}_n, \mathbf{Y}_{/n}\}$ )We then estimate posterior probabilities based on the training data using rBMS ($\hat{r}^{train}$) and MSA ($r'^{train}$). Finally, we calculate the log probability of the test data given the posterior probabilities estimates from the training data:

$$\begin{aligned} v_n^{BMS} &= \ln \int p(\mathbf{y}_n | \mathbf{r}) p(\mathbf{r} | \mathbf{Y}_{/n}, S_1) dr \\ &= \sum_{k=1}^{K} \hat{r}_k^{train} l_{nk}. \\ v_n^{MSA} &= \ln \int \int p(\mathbf{y}_n | \mathbf{r}) p(\mathbf{r} | \mathbf{Y}_{/n}, S) dr dS, \\ &= \sum_{k=1}^{K} r'^{train}_k l_{nk}. \end{aligned} \qquad (22)$$

(Here we assume that $S_1$ corresponds to the full model-space) These quantities can now be compared to assess the predictive performance of MSA and rBMS in any dataset.

**Results**

*Performance of model-space averaging with simulated data*

To assess the performance of MSA, when compared with rBMS, in situations where the ground truth is known, we simulated both behavioural data and fMRI time series, and performed model-based analyses as described above. In both cases, we simulated three different contexts. In the first, one model had zero probability of generating the data. In the second, all models contributed to the data. In the third, only one model contributed. By simulating these different contexts, we aim to explore a range of plausible scenarios, but this is not intended to be exhaustive.

Our simulation results suggest that, where one or more candidate models have zero probability, both model-space averaging and model-space selection provide significantly more accurate posterior probabilities than those derived from random-effects model selection, regardless of whether a greedy or exhaustive search is used (Figures 1 and 2). This difference resulted from the 'conservativism' of inference using a fixed model-space, as discussed above, which can be seen in the mean posterior probability estimates. Using MSA is particularly advantageous in the case where data were generated by a single model (Figure 2f,g Figure 3f,g) . By contrast, even when all candidate models contributed significantly to the data, rBMS produced posterior probability estimates that were only slightly more accurate (as assessed by the mean KL divergence between the true distribution and the estimated posteriors) than those calculated using MSA. This reflects the capacity of MSA to closely approximate rBMS where this is appropriate. Overall, the results of these simulations support the theoretical arguments given above which suggest that, overall, MSA tends to give more accurate results than rBMS, and is thus a generally preferable strategy for performing group-level model comparison.

In our simulations, both exhaustive and greedy search methods produced similar results, which is encouraging, given that there may be situations in which computational costs preclude the use of an exhaustive search over model-spaces. Given the limited range of situations we consider, however, this conclusion is necessarily a provisional one, and the safest strategy is to perform an exhaustive search wherever possible.

*Performance of model-space averaging in empirical data*

We first compared the posterior probabilities estimated by rBMS and MSA (using an exhaustive search) in two previously acquired datasets. In both cases, as expected, the results produced by rBMS showed evidence of a conservative bias, when compared with MSA (Table 1). (These differences were small in the case of dataset one, probably as a result of the small model-space employed). However, since we do not know the ground truth when using real data, simply examining the posteriors is not sufficient to provide evidence of greater accuracy when using MSA. To do this, we compared predictive accuracy of these two approaches using leave-one-out cross-validation on two previously acquired data sets (see *Description of datasets*). In both datasets predictive accuracy was consistently greater better for MSA than rBMS ($p < 0.0001$, Wilcoxon signed rank test in both cases), consistent with the theoretical and simulated results described above (Figure 4). This is important, because it demonstrates that MSA provides more accurate estimates of the posterior distribution over subject-level models than rBMS in real data, without making any assumptions about what the underlying distributions are. (More generally, cross-validation provides a tool for assessing, in any given dataset, whether MSA or rBMS is likely to provide more accurate estimates of the posterior probabilities)

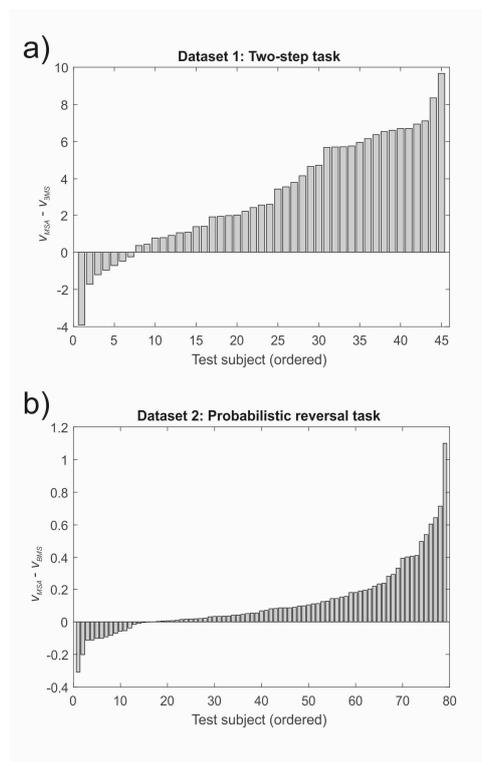

**Figure 4: Cross-validation results showing greater predictive accuracy for MSA than rBMS in two previously acquired datasets** [5,19]**. Values greater than zero indicate greater predictive accuracy for MSA than BMS for a particular subject's data, when the remaining data is used as a training set. In both cases, MSA produced considerably more accurate results.**

*Quantifying the evidence for population diversity*

We simulated two contexts (the 'diverse population' and 'inconclusive data' contexts), in which rBMS assigned similar posterior probabilities to two of the models ($m_1$ and $m_2$ in context one, and $m_3$ and $m_4$ in context two) (Figure 5a). It is thus not possible on this basis to decide whether the results in each context reflect a population containing two different agents, or simply the inadequacies of the data (and/or models). Comparison between model-spaces, however, clearly distinguishes between these contexts (Figure 5b). In context one, model-space $S_2$ is strongly supported, implying a diverse population. In context two, the greatest (and almost equal) probability is assigned to $S_6$ and $S_7$, which each contain only a single subject-level model, suggests that the data are inconclusive. We are thus able to formally distinguish a situation of genuine population diversity from one where no evidence for such diversity exists, even though the posterior probabilities over models are almost identical in both cases.

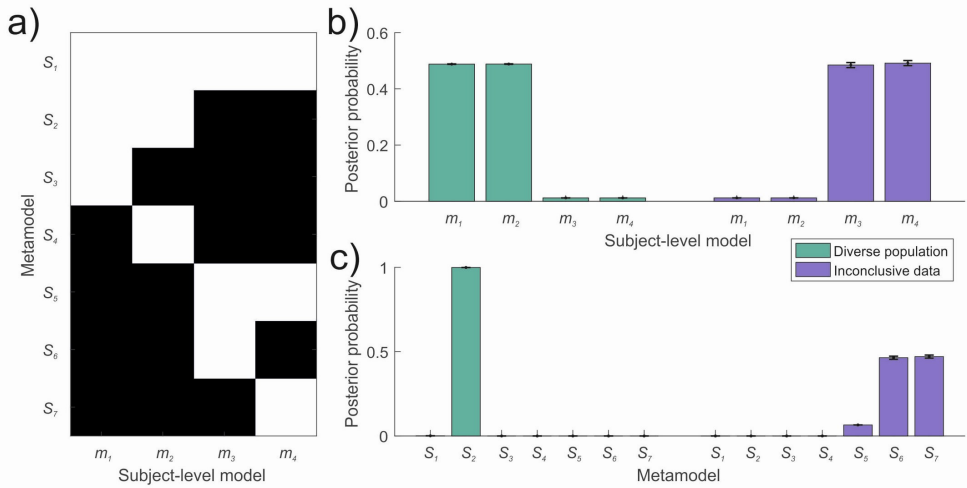

**Figure 5: Model-space selection dissociates genuine diversity within the experimental sample (green) from inconclusive data (purple). (5a is a graphical illustration of the models-spaces, with white indicating that a subject-level model is included within that model-space, and black that it is not. Note that we only consider a subset of the possible reduced model-spaces) In both contexts a posterior probability close to 0.5 is assigned to two different models ($m_1$ and $m_2$ in context one, $m_3$ and $m_4$ in context two) (5b). However, model-space comparison (5c) clearly favours a model-space containing both preferred subject-level models ($S_2$) in context one, but model-spaces containing only a single subject-level model ($S_6$ and $S_7$) in context two. (Error bars indicate 95% confidence intervals over 100 simulations)**

We also applied this approach to our previously acquired datasets, both of which used the BIC as an approximation to the log model evidence. For dataset 1 [19], both rBMS and MSA assign a significant posterior probability to both model-based ($m_1$) and hybrid ($m_3$) (model-based and model-free) learning schemes (Table 1). This raises the question of whether the data are best characterised as indicating a genuine mixture of learning strategies within the population studied, or if it is simply the case that model-based and hybrid models make predictions that are hard to disambiguate with these data. We tested this by comparing the evidence for three models-spaces, one containing both model-based and hybrid learners, and one containing each of these in isolation. The model-space containing both learners was assigned $P_1 \approx 1$, providing strong evidence for a diversity of learning strategies.

Similarly, in dataset 2 [5], both rBMS and MSA assign significant probability to two different models, $m_2$ and $m_3$ corresponding to inference over sequences of length two and three respectively (Table 1). Again, we compared three model-spaces, one containing both $m_2$ and $m_3$, and two containing only a single one of them. The model-space containing both learners was assigned $P_1 = 0.999$, which provides strong evidence for a variety of inference strategies in the population that we studied.

Taken together, these simulation and empirical results demonstrate the utility of model-space comparison as a means of formally testing for diversity within a population, something that we hope can be valuably employed by researchers in future studies.

*Comparing distributions of models between populations*

To illustrate the comparison of distributions of models between experimental populations, we first simulated behaviour on the probabilistic reversal task in two contexts, one in which the distributions over models differed between groups, and one in which they were identical. These could be clearly distinguished using the model-space comparison approach described in *Application 3: Comparing populations using model-space comparison* (Figure 6).

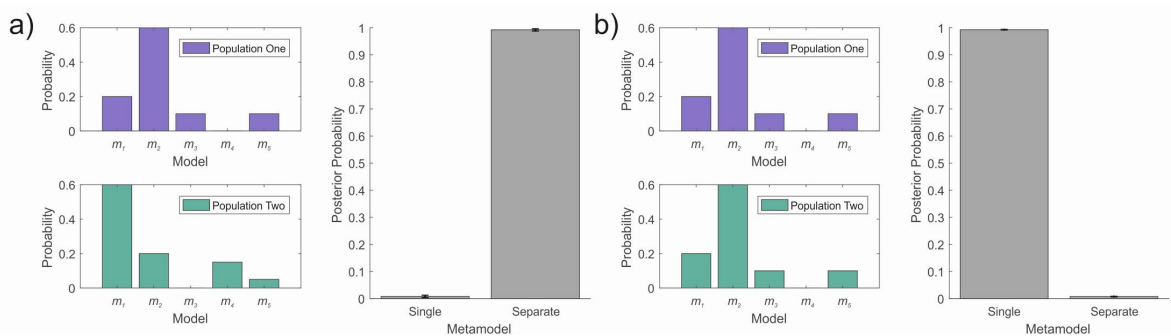

**Figure 6: Comparing the distribution of models between populations in simulated data. In context one (6a), the data were generated form two populations with separate model distributions (purple and green bars). In context two (6b), both populations had identical distributions. Model-space comparison clearly differentiates these two situations, correctly inferring the existence of separate distributions in context one and a single distribution in context two. (See main text for further details of simulations. Error bars indicate bootstrapped 95% confidence intervals)**

We also applied this approach to real data, specifically Dataset 2, which contains results from 43 younger adults, and 36 older adults. Previous work suggested a similar distribution of models in both groups [5], but we were unable to formally test this. As described above, model-space comparison provides a way of doing so, by either fitting a single metamodel, or two separate (but identically specified) metamodels, to the groups. Use of a single model-space is strongly preferred in this dataset, with a posterior probability of $P_0 = 0.991$, providing quantitative evidence for our previous informal conclusions.

**Discussion**

In this paper, we consider Bayesian inference over model-spaces, with a specific focus on model comparison in the context of multi-subject studies. This extends existing work that makes use of a single model-space [7,8,10], and allows pruning of redundant models, enhancing the accuracy of posterior probability estimates. We explore this theoretically, and show the effects of this using both simulated and empirical data. In particular, we compare inferential accuracy with random effects Bayesian model selection [7,8], the current state-of-the-art approach, and find significant improvementsThis advantage is likely to be particularly pronounced where the data are best explained using a relatively small subset of the full model space, as is likely to be the case in many real-world applications. (For example, one might imagine that the connectivity structure of functional brain networks is likely to be highly similar between subjects) When using the variational implementation we propose here, model-space averaging requires only very limited additional computation when compared with rBMS, and thus provides an attractive and straightforward method for achieving more accurate results during statistical analysis.

Inference over model-spaces can also be used to formally address a number of questions about the distribution of models in one or more populations, in a way that is impossible with existing model comparison procedures (though see [7]). In this paper, we have used model-space comparison to test to generate novel results that supplement our previously reported findings [5,19]. In particular, we provide evidence that subjects adopt a genuine diversity of strategies when performing both the two-step [19] and probabilistic reversal [5] tasks. In addition, we illustrate the use of model comparison to formally test the hypothesis that a similar distribution of strategies is found in both older and younger subjects [5]. These results represent significant extensions of what was previously reported, as well as illustrating the more general point that model-space comparison provides an important tool for formally addressing questions about the distribution of models within a population.

A number of other applications of model-space comparison are undoubtedly conceivable, and we expect that these will be explored in future studies. For example, one natural extension of the work presented here is to include inference over families of models [12], however this may require the use of a sampling scheme, as discussed in [12], which we will explore in future work. Previous work proposing the use of protected exceedance probabilities [7] can also be understood within the broader framework of model-space comparison, since it rests on a comparison between the full metamodel and a 'null' metamodel, in which all models are constrained to have the same probability.

In general, it seems likely that between-subject differences in brain and behaviour, including those that manifest themselves as psychopathology [20–22], will at times be manifest as differences in the types of model that best explain experimental data [23], rather than simply different parameter values within similar models [24]. This might correspond, for example, to the tendency to adopt fundamentally different cognitive strategies to solve a particular task [6,19,25], or neuronal networks structured in fundamentally different ways. Model-space comparison provides a novel way to address questions at this level, and thus represents a useful tool for possible future studies.

Model-space averaging helps to reduce the inherent conservativism of rBMS, and more generally, by reducing the diluting effect of suboptimal models on the posterior probabilities assigned to more appropriate models [1,2], it serves to mitigate the negative effects of large model spaces. This is important, as it reduces the dependence of the results of an analysis on whether *prima facie* plausible but poorly fitting models are included within the model-space or not[2]. However, this does not solve the problem of generating appropriate subject-level models in the first place (as opposed, for example, to Bayesian model reduction [26,27]).

Recently, there has been considerable interest in the neuroimaging community in using parametric empirical Bayes (PEB) rather than rBMS to analyse multi-subject dynamic causal modelling data [27]. This is an appealing approach for models that have a nested structure, such as DCMs, but is unlikely to be useful in situations where models have fundamentally different structures. (For example, when considering model-based and model-free reinforcement learning agents [6,19,28], or when comparing Bayesian and reinforcement learning agents [5,29])

An exciting extension of rBMS has been proposed, in which hierarchical parameter estimation for subject-level models is performed simultaneously with inference over models [10]. (This can be contrasted with the approach discussed here and elsewhere [2,7,8], in which fitting of subject-level models and inference over models is performed sequentially) This approach could, in principle, be combined with model-space averaging as described here, provided the inference scheme being employed provides an accurate estimate of the evidence for each model-space, something we intend to explore in future work.

An alternative way to induce sparsity over models is to assign very low (close to zero) values to $c$, which governs the prior concentration parameters $\boldsymbol{\alpha}_{\bullet j}^{prior}$ for all models included within the current model-space. We have not discussed this here, for two reasons. First, we found in simulations that are not reported here, that the results produced by averaging over different values of $c$ were generally less accurate than those produced by using reduced model-spaces. Second, we have previously found that the variational approximation used here and elsewhere [7,8] becomes inaccurate for low values of $c$ [2]. We thus postpone consideration of this topic until future work, in which we will consider model-space inference using sampling rather than variational inference. In this paper we propose a new approach to inferring on which models best explain a particular data set. This extends existing work that makes use of a single model-space [7,8,10], by performing inference over model-spaces. Importantly, and unlike in existing approaches, this permits consideration of situations in which zero probability is assigned to one or more candidate models. We show, using simulated behavioural and fMRI data, that in most plausible contexts, this results in more accurate posterior probability estimates than that furnished by random effects model selection, which constitutes the standard approach in the field [7,8]. The methods we propose here thus represent a significant step forward in developing more accurate tools for Bayesian inference in multi-subject studies. Additionally, model-space comparison provides a formal framework for addressing hypotheses about distributions of models, which is likely to be important for future studies exploring population diversity at this level.

## Acknowledgements

We thank R. Dolan, S-C. Li and J. Goslin for their contributions to the datasets we analyse here. THBF is supported by a European Research Council (ERC) Starting Grant under the Horizon 2020 program (Grant Agreement 804701)